\documentclass{PoS}
\usepackage{amsmath}
\usepackage{epsfig}
\usepackage{graphicx}
\usepackage{multirow}
\usepackage{multicol}
\usepackage{comment}
\title{Charmonium spectroscopy from an anisotropic lattice study}

\ShortTitle{Charmonium spectroscopy from an anisotropic lattice study}

\author{\speaker{Liuming Liu}, \speaker{Sin\'ead M. Ryan}, Mike Peardon, Graham Moir and Pol Vilaseca\\
        School of Mathematics, Trinity College Dublin, Dublin, Ireland. \\
        E-mail: \email{liuming@maths.tcd.ie}}

\abstract{We present a progress report on our study of the charmonium spectrum in full QCD on anisotropic lattices generated by the Hadron Spectrum 
Collaboration. We adopt a large basis of interpolating operators to extract the excited charmonium states using the
variational method. A detailed spectrum of excited charmonium mesons in many $J^{PC}$ channels is obtained. Hybrid states with 
exotic and non-exotic quantum numbers are determined and preliminary results from a study of disconnected contributions to the 
$\eta_c$ are presented.}

\FullConference{The \uppercase\expandafter{\romannumeral29} International Symposium on Lattice Field Theory\\
         July 11-16 2011\\
         Squaw Valley, Lake Tahoe\\
         CA, USA}

\begin{document}

%\maketitle

\section{Introduction}
In recent years there has been a resurgence of interest in charmonium spectroscopy, mainly due to the unexpected, narrow 
charmonium states discovered by BaBar, Belle and CLEO collaborations. The properties of these new states, called 
$X$, $Y$ and $Z$, have been discussed in some recent reviews ~\cite{Brambilla:2011, Nielsen:2010}.  With the experiments at LHC and BES-\uppercase\expandafter{\romannumeral3} and the planned PANDA experiment at the new FAIR accelerator, more information about these states and the 
discovery of further states can be expected. Theoretical predictions, especially those from lattice QCD
calculations, will provide important guidance for these experimental efforts. 

Recent lattice studies of charm spectroscopy include ~\cite{Dudek:2008ce, Bali:2011cds} and progress was reported by a number of 
groups at this conference~\cite{Bali:2011cs, Mohler:2011, Namekawa:2011cx}. 
In this work, we study the charmonium spectrum, including radially and orbitally excited conventional
$c\bar{c}$ states and hybrids, on an anisotropic lattice ensemble generated by the Hadron Spectrum
Collaboration~\cite{Lin:2008pr}. We also study the contribution of disconnected diagrams. The distillation quark-field smearing algorithm~\cite{Peardon:2009gh}, which is essential to improve the precision and enable efficient calculations of a broad range of hadron correlation functions, is used for the first time in charmonium spectrum study. By the application of the variational method~\cite{Luscher:1990va} with  a large basis of derivative-based interpolating operators, we are able to extract and identify high spin (up to spin-4) states as well as exotic and non-exotic hybrids. 

 This paper is organized as follows: we first give the details of the lattice setup and the tuning of parameters in the quark action, then explain the construction of the interpolating operators and the variational
method as well as the method used to identify the extracted states. 
In section~\ref{Sec:Results} our results are presented by $J^{PC}$ while section~\ref{Sec:Summary} summarizes our calculation and describes ongoing work.

\subsection{Lattice Details}
To study high-spin states, exotic states and radial excitations the anisotropic lattice has proven very 
effective~\cite{Dudek:2010em}. By making the lattice spacing in the temporal direction, $a_t$ finer than the 
spatial, $a_s$ additional resolution is gained in the direction in which measurements are made for resonable computational cost. In this work we use the lattices and parameters of the Hadron Spectrum Collaboration which are described in detail in 
Ref.~\cite{Lin:2008pr}. In brief, we use a Symanzik-improved gauge action and a tree-level tadpole-improved 
Sheikholeslami-Wohlert fermion action with stout-smeared~\cite{Morningstar:2003gk} spatial links. The anisotropy 
$\xi=a_s/a_t=3.5$ and 
the charm quark is simulated with the same (relativistic) action as used for the light quarks. 
Distillation~\cite{Peardon:2009gh} 
is useful in this study since it improves the signal-to-noise in traditionally noisy states such as hybrids and disconnected 
correlators and simplifies correlator constuction allowing for a large basis of operators with inversions on 
all timeslices. Some relevant parameters are summarised in Table~\ref{tab:latt_details}. 
\begin{table}[h]
\begin{center}
\begin{tabular}{ccccccc}
$N_f$ & Volume         & $N_{\rm cfgs}$ & $m_\pi$ (MeV) & $\xi$ & $N_{\rm ev}$  & $a_t^{-1}(\Omega)$ (GeV) \\
\hline
2+1 & $16^3\times 128$ & 96           & 396           & 3.5   & 64 & 5.667 
\end{tabular}
\end{center}
\caption{\label{tab:latt_details} 
The reader is referred to Ref.~\cite{Lin:2008pr} for more details. 
In this study we use 64 distillation eigenvectors and inversion on all timeslices ($N_t=128$) 
in the distillation space.}
\end{table}

\subsection{Heavy quarks on an anisotropic lattice}
\begin{figure}[h]
\centering
\includegraphics*[width=0.45\textwidth]{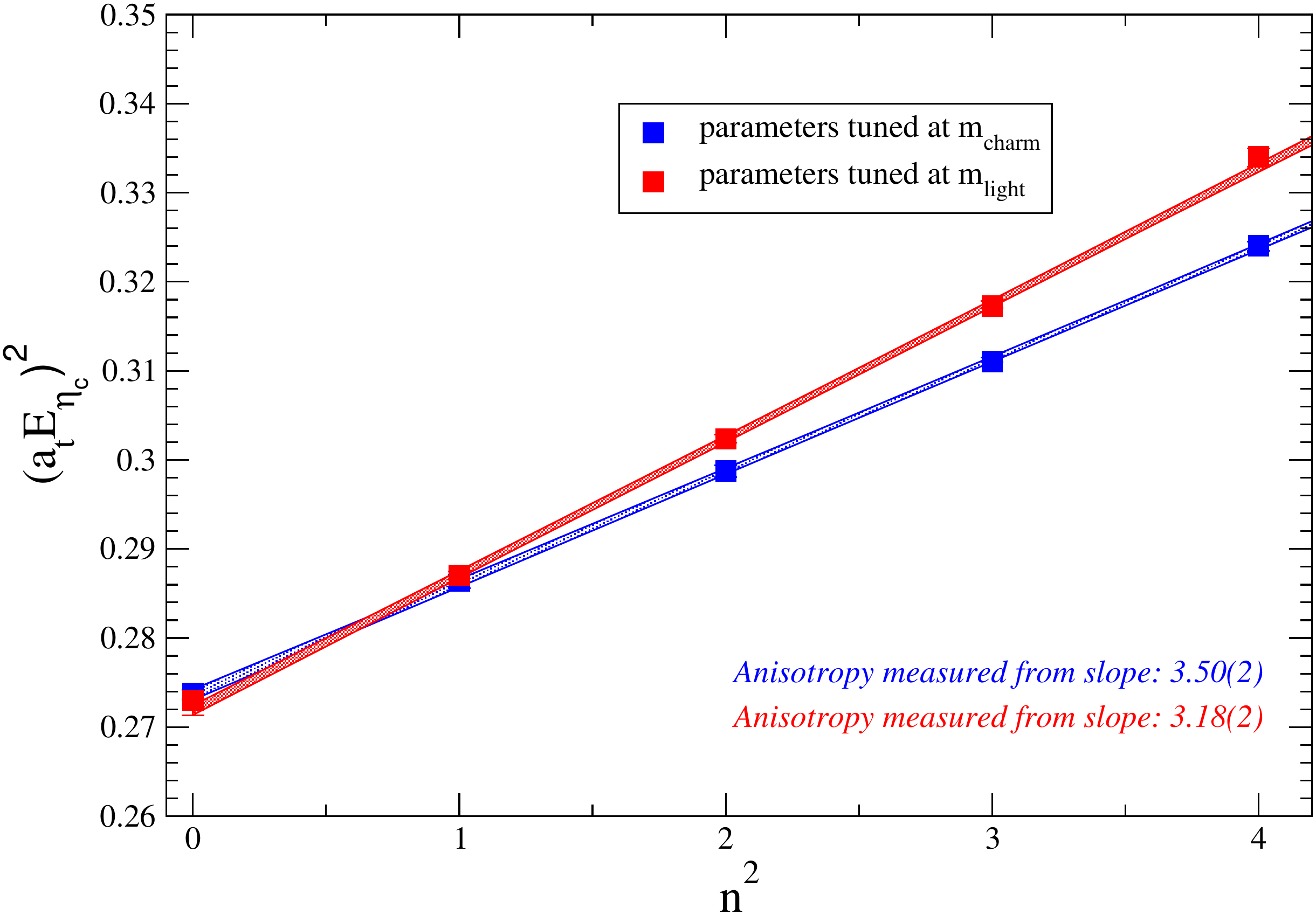}
\includegraphics*[width=0.45\textwidth]{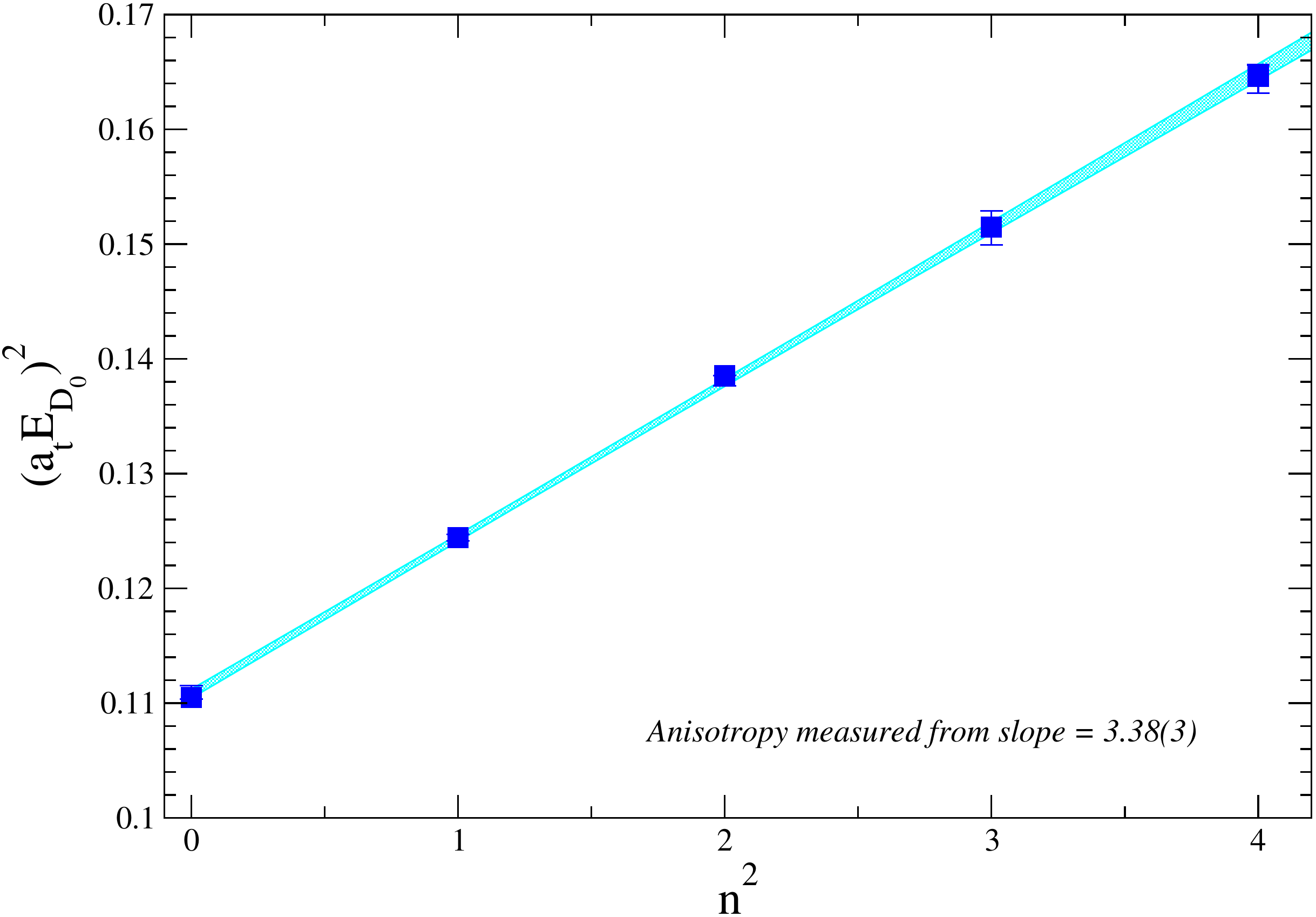}
\caption{\label{fig:dispersions} The dispersion relations for the $\eta_c$ (left plot) and $D$ (right plot) 
mesons. The left plot shows two dispersion relations: the points in red are determined using 
anisotropy parameters tuned at light quark masses while the blue points show the 
dispersion relation determined after a retuning of the fermion anisotropy at the charm quark mass. 
The measured anisotropies are also shown, determined from the slope by a linear fit including all points out to $n^2=4$. 
The right plot shows the dispersion relation for the $D$ meson using the parameters tuned 
at $m_{\rm charm}$, on a subset of 15 configurations. 
The tuning of the anisotropy at $m_{\rm charm}$ also results in a small mass retuning 
which has been taken into account in these plots.} 
\end{figure}

In dynamical simulations on anistropic lattices the bare anisotropy must be tuned, in the gauge and 
fermion sectors simultaneously, to ensure that the measured quantity in a physical probe takes its desired value; 
3.5 in this case. 
This nonperturbative tuning by the Hadron Spectrum Collaboration is described in Ref.~\cite{Edwards:2008ja} where the dispersion relation for light quark masses was used to tune the anisotropy in the fermion action. We have 
investigated the accuracy of this bare parameter, tuned at $m_{\rm light}$ for charmonium (and $D$) physics and find 
that the parameter tuning in the light sector leads to an underestimate of the slope in the charmonium dispersion relation of $\sim$15\%. 
Figure~\ref{fig:dispersions} shows two different dispersion relations: for the $\eta_c$ on the left and the $D$ 
meson on the right. The left plot compares the dispersion relations of the $\eta_c$ when the heavy quark 
perambulators are calculated with a bare anisotropy tuned at $m_{\rm light}$ and at $m_{\rm charm}$. 

The measured values are shown on the plot, determined from 
$\displaystyle{ {\gamma_f}=\frac{\pi}{8\sqrt{b}}}$, where $b$ is the slope of the dispersion relation.  
The ``retuned'' bare anisotropy is 
then used to determine the dispersion relation for the $D$ meson. We find the anisotropy measured from the slope 
is 3.38(3), in reasonable agreement with the target of 3.5. In passing we also note that 
the mass of the $D$ meson is determined to be $1883(64)$ MeV (for unphysically-heavy light quarks). 
The measured anisotropy 
is precisely determined in both cases, with a statistical error less than a percent and 
excellent fits out to momenta of $n^2=4$. A more detailed analysis of the heavy-light 
meson spectrum is underway but some preliminary results on just 15 configurations, 
include the mass difference $\Delta(M_{D^\ast}-M_D)$=137(4) MeV to be compared to the 
experimental result of 142.12(7)MeV. 

\subsection{Disconnected diagrams}
In this study we present our preliminary results for the disconnected contributions to the $\eta_c$, on 
a subset of 39 configurations. The $\eta_c$ is a colour singlet and Wick contractions produce bubble 
diagrams, as well as the familiar connected diagrams, which should be included for an accurate 
determination of the mass. It can be argued that in charmonium these disconnected diagrams are 
OZI suppressed, however to achieve a precision on mass determinations at the percent level it is valuable to 
determine their contribution more reliably.

\begin{figure}[h]
\centering
\includegraphics*[width=0.7\textwidth]{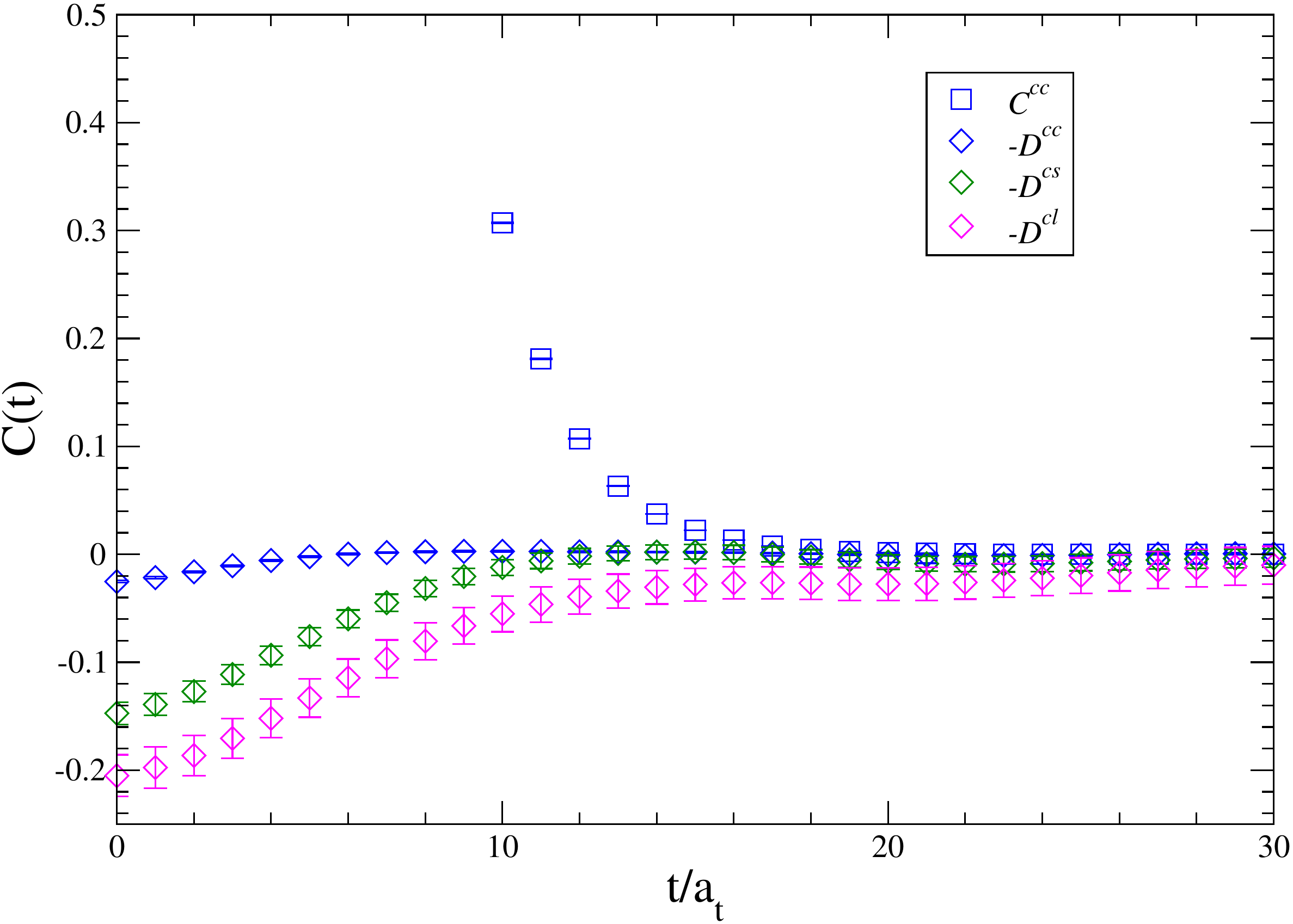}
\caption{\label{fig:disco_all}The disconnected contributions to the $\eta_c$ from the charm, strange and light sectors. The 
connected correlator is also shown. The disconnected correlators are determined in this preliminary work on just 39 
configurations but nevertheless a clear signal is determined which persists over a number of timeslices. We are currently 
increasing statistics for a full analysis.}
\end{figure}

Distillation simplifies the calculation of disconnected diagrams considerably. 
The correlation function becomes a product of traces over rank $M$ matrices, where 
$M$ is the number of distillation eigenvectors used in the calculation. 
\begin{equation}
C^{\rm disc}(t^\prime,t)=\rm{Tr}\left[\Phi^A(t)\tau(t,t)\right]
                     \rm{Tr}\left[\Phi^B(t^\prime)\tau^\dagger(t^\prime,t^\prime)\right],
\end{equation}
where $\tau(t,t)$ are the perambulators encoding the quark dynamics and the $\Phi$ contain information about the 
spin structure of the meson. 
In Figure~\ref{fig:disco_all} we show the disconnected correlators which are relevant 
for the $\eta_c$ as well as the connected correlator. 
Note that clear signals for the disconnected contributions from the charm, strange and light sectors all persist 
over a reasonable number of timeslices. One can write a matrix in flavour space to describe the mixing of these 
contributions and subsequently extract the full $\eta_c$ correlator including all disconnected effects as well as 
the mixing angles, and this is ongoing work. 
\subsection{Construction of interpolating operators}
The hadron spectral information is extracted from the time dependence of two-point correlation functions between 
creation and annihilation operators. The simplest meson interpolating operators are the local fermion bilinears 
$\bar{\psi}(x)\Gamma\psi(x)$, where the quantum numbers are determined by the choice of gamma matrix, $\Gamma$. 
However, these local operators allow for only a limited set of quantum numbers and to study higher spin 
and exotic states one must use non-local operators. In this work, 
we use the derivative-based operators~\cite{Dudek:2008ce, Dudek:2010em, Liao:2002ch} which are constructed by 
using spatially-directed gauge-invariant derivatives within a fermion bilinear. The general form is 
\begin{equation}
\bar{\psi}\Gamma \overleftrightarrow{D}_i \overleftrightarrow{D}_j \cdots\psi ,
\end{equation}
where $\overleftrightarrow{D} \equiv \overleftarrow{D} - \overrightarrow{D}$. The use of the 
``forward-backward'' derivative is helpful to ensure that the operators have definite charge-conjugation 
at finite momentum, but not necessary 
for zero momentum which is the only case we consider in this work. It is straightforward to build this type of operator with definite spin, parity and charge-conjugation in continuum space. The derivatives can be expressed in a basis transforming like 
spin-1:
\begin{equation}
\overleftrightarrow{D}_{m=-1} = \frac{1}{\sqrt{2}}(\overleftrightarrow{D}_x - i \overleftrightarrow{D}_y), \quad 
\overleftrightarrow{D}_{m=0} = i \overleftrightarrow{D}_z, \quad 
\overleftrightarrow{D}_{m=1} = -\frac{1}{\sqrt{2}}(\overleftrightarrow{D}_x + i \overleftrightarrow{D}_y). 
\end{equation} 
Using the standard $SO(3)$ Clebsch-Gordan coefficients, one can combine any number of derivatives with gamma matrices to construct operators with the desired quantum numbers. The choice of gamma matrix plays a role in setting the parity and charge-conjugation and the relevant matrices for this study are listed in Table~\ref{Table:GammaMatrix} with a naming scheme, as used in 
Ref.~\cite{Dudek:2008ce}.
\begin{table}[h]
\begin{center}
\begin{tabular}{c|cccccccc}
         &$a_0$ &$\pi$ &$\pi_2$ &$b_0$ &$\rho$ &$\rho_2$ &$a_1$ &$b_1$ \\
\hline
$\Gamma$ &1 &$\gamma_5$ &$\gamma_0 \gamma_5$ &$\gamma_0$ &$\gamma_i$ &$\gamma_i \gamma_0$ &$\gamma_5 \gamma_i$ 
         &$\gamma_i \gamma_j$ 
\end{tabular}
\end{center}
\caption{\label{Table:GammaMatrix} Gamma matrix naming scheme.}
\end{table}
With one derivative, the operators of spin $(J, M)$ can be formed as
\begin{equation}
(\Gamma \times D^{[1]}_{J=1} ) ^{J, M} = \sum_{m_{\Gamma}, m_2} \langle J_{\Gamma}, m_{\Gamma}; 1, m_2 | J, M \rangle 
\bar{\psi}\Gamma_{m_{\Gamma}} \overleftrightarrow{D}_{m_2} \psi ,
\end{equation} 
where $J$ is the continuum spin and $M$ is the number of components of that spin. 
For scalar-like gamma matrices, $J_{\Gamma} = 0, m_{\Gamma} = 0$. For vector-like gamma matrices, $J_{\Gamma} = 1, 
m_{\Gamma} = -1, 0, 1$. 
Two derivatives can be combined as a definite spin $J_D$, and then combined with the gamma matrix as
\begin{equation}
(\Gamma \times D^{[2]}_{J_D} ) ^{J, M} = \sum_{m_{\Gamma}, m_D, m_1, m_2} \langle J_{\Gamma}, m_{\Gamma}; J_D, m_D | J, M 
\rangle  \langle 1, m_1; 1, m_2 | J_D, M_D \rangle  \bar{\psi}\Gamma_{m_{\Gamma}} \overleftrightarrow{D}_{m_1} 
\overleftrightarrow{D}_{m_2} \psi.
\end{equation} 
At the three-derivative level, we follow the convention in Ref.~\cite{Dudek:2010em} for the ordering in which we 
couple the derivatives. The first and third derivatives are coupled to a definite spin $J_{13}$  and then the second derivative is coupled with $J_{13}$ to form a definite spin $J_D$. 
\begin{align}
(\Gamma \times D^{[3]}_{J_{13}, J_D} ) ^{J, M} =  \sum_{m_\Gamma, m_D, m_{13}, m_1, m_2, m_3}& \langle J_{\Gamma}, m_{\Gamma}; J_D, m_D | J, M \rangle  \langle 1, m_2; J_{13}, m_{13} | J_D, M_D \rangle  \\ \nonumber
& \times \langle 1, m_1; 1, m_3 | J_{13}, M_{13} \rangle  \bar{\psi}\Gamma_{m_{\Gamma}} \overleftrightarrow{D}_{m_1} \overleftrightarrow{D}_{m_2} \overleftrightarrow{D}_{m_3} \psi.
\end{align} 
The operators built in this simple way have definite charge-conjugation. This is possible because the 
forward-backward derivatives are used. 
This procedure can be extended to any number of derivatives. In this work, we use operators with all 
combinations of up to three derivatives. 

In lattice calculations, states are not classified according to the spin (J, M), but rather according to the irreducible representations (irreps) of the cubic group. Different continuum spin components can be distributed across several irreps. Table~\ref{Table:Irreps} shows the five irreps, $\Lambda$ of the cubic group and their continuum spin content. 
\begin{table}[h]
\begin{center}
\begin{tabular}[width=\textwidth]{cccccc}
\hline
 Irreps ($\Lambda$) &$A_1$ &$A_2$ &$T_1$ &$T_2$ &$E$ \\
 $d_\Lambda$ & 1 &1 &3 &3 &2 \\
 $J$ &0, 4, 6, $\cdots$ & 3, 6 ,7, $\cdots$   & 1, 3, 4, $\cdots$ & 2, 3, 4, $\cdots$ &2, 4, 5, $\cdots$ \\
 \hline
 \end{tabular}
 \end{center}
 \caption{\label{Table:Irreps} The table shows the irreducible representations $\Lambda$ of the cubic group,
 together with their dimensions ($d_\Lambda$) and continuum spin content $J$.}
 \end{table}
The continuum operators described above must then be subduced into lattice 
irreps. The subduction can be performed using linear combinations of the $M$ components for each $J$: 
\begin{equation}\label{Eq:Subduction}
\mathcal{O}^{[J]}_{\Lambda, \lambda} \equiv \sum_M S^{J, M}_{\Lambda, \lambda} \mathcal{O}^{J, M}, 
\end{equation}
where $\lambda = 1, \cdots, d_\Lambda$, is the "row" of the irrep $\Lambda$, $\mathcal{O}^{J, M}$ is the continuum 
operator of spin $(J, M)$ and $S_{\Lambda, \lambda}^{J, M}$ are the subduction coefficients which form an orthogonal matrix $\sum_M S_{\Lambda, \lambda}^{J, M} S_{\Lambda^\prime, \lambda^\prime}^{J, M} = \delta_{\Lambda, \Lambda^\prime} 
\delta_{\lambda, \lambda^\prime}$ . For the details of the derivation of the subduction coefficients see Ref.~\cite{Dudek:2010em}. 
\subsection{Variational method}
The variational method~\cite{Luscher:1990va} is a very powerful tool to extract multiple excited states, utilizing a large basis of operators within a given symmetry channel. Mathematically this method involves solving the generalized eigenvalue problem
\begin{equation}
C(t) v_a = \lambda_a(t) C(t_0) v_a,
\end{equation}
where $C(t)$ is the correlator matrix at time slice $t$, $C_{ij}(t) = \langle \mathcal{O}_i (t)\bar{\mathcal{O}}_j(0)\rangle$. The eigenvectors are orthogonal on the metric $C(t_0)$, $v_a^\dagger C(t_0)v_b = \delta_{a,b}$. 
The eigenvalues $\lambda_a(t)$ satisfy
\begin{equation}\label{Eq:PrinCorr}
\lambda_a(t) = e^{-m_a(t-t_0)}(1+\mathcal{O}(e^{-\Delta m_a(t-t_0)}),
\end{equation}
where $m_a$ is the mass of a state labeled by $a$ and $\Delta m_a$ is the distance of $m_a$ from the mass of the nearest state. At large $t$, the principal correlators tend to $\sim e^{-m_a(t-t_0)}$. In practice, a second exponential is used in the fit form: 
\begin{equation}
\lambda_a(t) =(1-A_n) e^{-m_a(t-t_0)} + A_n e^{-m_a^\prime(t-t_0)},
\end{equation} 
where $A_n$, $m_a$ and $m_a^\prime$ are fitting parameters. 

Choosing a proper $t_0$ plays an important role because we use a finite number of operators. The eigenvectors are 
forced to be orthogonal on the metric $C(t_0)$ by the solution procedure. This will only be a good approximation 
to the true orthogonality if the correlator at $t_0$ is dominated by the lightest dim($C$) states. In this work, 
we follow the ``reconstruction'' procedure described in Ref.~\cite{Dudek:2008ce} to choose $t_0$: the correlator matrix is reconstructed from the extracted masses and the eigenvectors, a $\chi^2$-like quantity is defined and 
$t_0$ is chosen so that this is minimised. 

\subsection{Spin identification}
As the lattice spacing approaches zero, the rotational symmetry is restored and the spins can be identified 
by the emergence of energy degeneracies between different irreps in the continuum limit. For example, a spin-2 state would be expected to appear as degenerate energies within the $T_2$ and $E$ irreps. However, this method requires repeating the calculation 
at a series of different lattice spacings. In this work we only have a single lattice spacing. 

\begin{figure}[h]
 \begin{center}
 \includegraphics*[width = 0.7\textwidth]{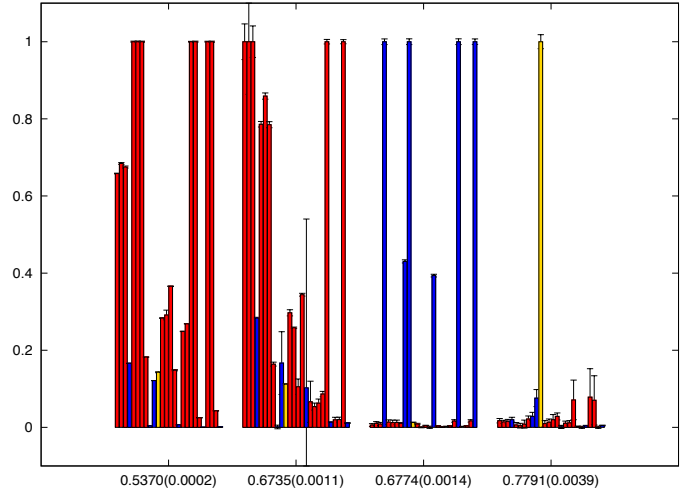}
 \end{center}
 \caption{\label{Fig:overlap} The overlaps of four lowest-lying states onto the 26 different operators used in the $T_1^{--}$ 
analysis. The color-coding indicates the continuum spin $J$ from which the operators are subduced 
(red: J=1, blue: J=3, yellow: J=4). }
 \end{figure}

An alternative approach valid for finite lattice spacing is to consider the "overlap" of the state extracted in the lattice 
irreps onto the operators, 
$\langle 0 |\mathcal{O} | n \rangle$. From Eq.~\ref{Eq:Subduction} one can see that the operators constructed on 
the lattice carry a "memory" of the continuum spin from which it is subduced. If our lattice is reasonably close to the continuum limit, we can expect that an operator subduced from spin $J$ only only overlap strongly onto those states of continuum spin $J$. Figure~\ref{Fig:overlap} shows the overlap histograms of four low-lying states in $T_1^{--}$ channel. 
It is clear that the first two states overlap strongly onto $J=1$ operators, the second state overlaps onto 
$J=3$ operators and the fourth state overlaps onto $J=4$ operators. 

To be more quantitative, we can compare the overlaps obtained in different irreps. The continuum operators have 
definite spin $\langle 0 | \mathcal{O}^{J, M} | J^\prime, M^\prime \rangle = Z^{[J]} \delta_{J, J^\prime} \delta_{M, M^\prime}$. 
Therefore, as continuum symmetries are restored lattice operators should obey 
$\langle 0 | \mathcal{O}^{[J]}_{\Lambda, \lambda} | J^\prime, M^\prime 
\rangle = S^{J, M^\prime}_{\Lambda, \lambda}Z^{[J]} \delta_{J, J^\prime}$. The value of $Z^{[J]}$ is common in different irreps. This suggests that we can identify a state by comparing the Z-values obtained in different irreps. For example, a $J=2$ state created by $J=2$ operator will have the same Z-value in each of the $T_2$ and $E$ irreps. 
\section{Spectrum Results}\label{Sec:Results}
The masses of charmonium states are calculated in every $\Lambda^{PC}$ channel. Continuum spins are assigned using the method described above. In the following we will discuss each $PC$ combination separately. The continuum spins are consistently colour-coded in the discussion that follows: black for $J=0$, red for $J=1$, green for $J=2$, blue for $J=3$ and yellow for $J=4$. We use brown when we 
cannot make an unambiguous spin assignment. Figures~\ref{Fig:Jpp} to ~\ref{Fig:Jmp} show our results in each $PC$ combination, ordered by lattice irrep and colour-coded by continuum spin, $J$. The lowest thresholds for open-charm, $D\bar{D}$ and $D_s\bar{D}_s$ are also shown. These thresholds have been calculated on the same lattices, described in Table~\ref{tab:latt_details}.
\subsection{$J^{++}$}
\begin{figure}[h]
\centering
\includegraphics*[width=0.7\textwidth]{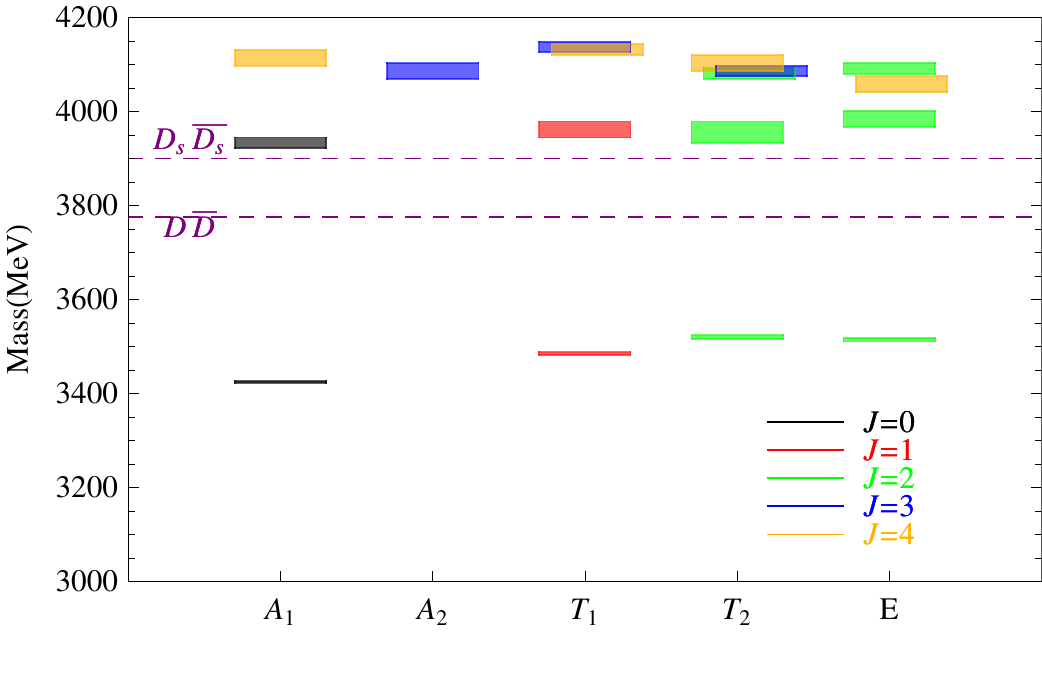}
\caption{\label{Fig:Jpp} The plot shows the masses of the extracted $J^{++}$ states by lattice irreps.}
\end{figure}

Figure~\ref{Fig:Jpp} shows the extracted $J^{++}$ states in lattice irreps. 
%%%
The lowest band of states, located in the irreps $A_1$, $T_1$, $T_2$ and $E$, are identified as the 
near-degenerate P-wave triplet $0^{++}$, $1^{++}$ and $2^{++}$ by looking the overlaps of these states with the 
operators in each channel. The assignment of $2^{++}$ is also supported by comparing the overlaps onto the 
operator $(\rho \times D^{[1]}_{J=1} )^{J=2}$ in the $T_2$ and $E$ channels: 

\begin{equation}
\frac{Z_{E}(\text{ground state})}{Z_{T_2}(\text{ground state})} = 0.997(4).
\end{equation}
This result clearly indicates that the ground states in $T_2$ and $E$ are different components of a $2^{++}$ state. 
The second band of states, at around 3950~MeV, are also located in the irreps $A_1$, $T_1$, $T_2$ and $E$.
At this energy range they could all belong to a single $4^{++}$ state considering the continuum spin contents of
these four irreps. However, the overlap histograms indicate that they are likely to be $0^{++}$, $1^{++}$ and $2^{++}$ states. The $2^{++}$ assignment is also supported by comparing the overlaps onto the operator $(\rho \times
D^{[1]}_{J=1} )^{J=2}$ in $T_2$ and $E$ channel: 
\begin{equation}
\frac{Z_{E}(\text{1st exited state})}{Z_{T_2}(\text{1st exited state})}\big((\rho \times D^{[1]}_{J=1}
)^{J=2}\big ) = 0.92(8).
\end{equation}
Another issue arising here is that this $2^{++}$ state could be either a radially exited P-wave or an F-wave.
Checking the overlaps of this state onto an F-wave operator $(\rho \times D^{[3]}_{J_{13}=2, J=3} )^{J=2}$ in $T_2$
and $E$, we find they are small in comparison to the overlaps onto the operator $(\rho \times D^{[1]}_{J=1}
)^{J=2}$. Thus, identifying this $2^{++}$ state as a radially exited P-wave is preferable. Therefore, 
we can identify these near-degenerate $0^{++}$, $1^{++}$, $2^{++}$ states as the radially exited P-wave triplet.

The remaining state in the $A_1$ channel can be identified as a $4^{++}$ state by its large overlap on the $J=4$ 
operator $(\rho \times D^{[3]}_{J_{13}=2, J=3} )^{J=4}$. The only state in $A_2$ is probably a $3^{++}$. The remaining two states in $T_1$ are expected to be a $3^{++}$ and a $4^{++}$. We can assign the second exited state as a $3^{++}$ and the third exited state as a $4^{++}$ by simply looking the overlap histogram in this channel. Similarly, the remaining three states in $T_2$ channel are identified as a $2^{++}$, a $3^{++}$ and a $4^{++}$. The remaining two states in $E$ are identified as a $2^{++}$ and a $4^{++}$. These assignments are all well supported by comparing the overlaps between different irreps. Comparing the overlaps of the assigned $2^{++}$ states in $T_2$ and $E$ onto the operator $(\rho \times D^{[3]}_{J_{13}=2, J=3} )^{J=2}$, we get the ratio to be $1.12(5)$. Comparing the overlaps of the assigned $3^{++}$ states in $A_2$, $T_1$ and $T_2$ channels onto the operator $(\rho \times D^{[3]}_{J_{13}=2, J=3} )^{J=3}$ , we have
\begin{equation}
\frac{Z_{A_2}(\text{ground state})}{Z_{T_1}(\text{2nd exited state})} = 0.94(2), \quad 
\frac{Z_{A_2}(\text{ground state})}{Z_{T_2}(\text{3nd exited state})} = 0.99(2). 
\end{equation}
For the $4^{++}$ state distributed in $A_1$, $T_1$, $T_2$ and $E$ channels, the overlaps onto the operator 
$(\rho \times D^{[3]}_{J_{13}=2, J=3} )^{J=4}$ all appear to be compatible with a common state assignment: 
\begin{equation}
\frac{Z_{A_1}(\text{2nd exited state})}{Z_{T_1}(\text{3rd exited state})} = 0.97(1),  \frac{Z_{A_1}(\text{2nd exited  
state})}{Z_{T_2}(\text{4th exited state})} = 0.98(2),   \frac{Z_{A_1}(\text{2nd exited state})}{Z_{E}(\text{2nd exited state})} = 1.09(3).
\end{equation}
\subsection{$J^{--}$}
\begin{figure}[h]
\centering
\includegraphics*[width=0.7\textwidth]{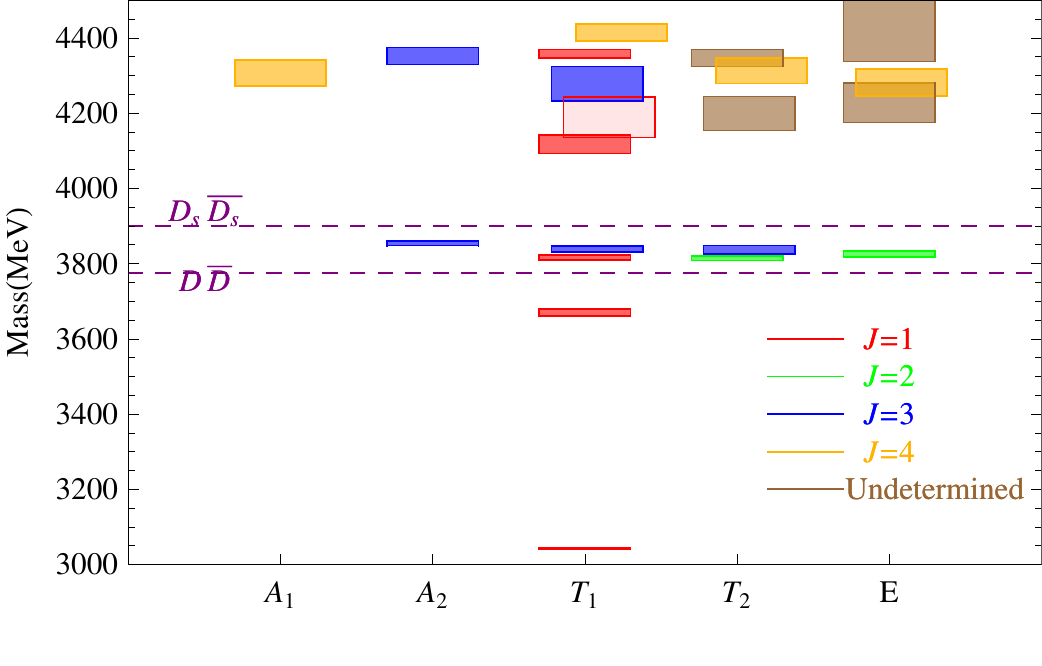}
\caption{\label{Fig:Jmm} The plot shows the masses of the extracted $J^{--}$ states by lattice irreps. }
\end{figure}

Figure~\ref{Fig:Jmm} shows the extracted $J^{--}$ states organized by lattice irreps. 
The lowest state in the $T_1$ channel can be identified as the $J/\psi$. The first excited state in $T_1$ is also a $1^{--}$ by looking the overlap histogram, which is very similar to the histogram for the ground state. Thus, we identify it as a radial exitation of the ground state, $J/\psi(2S)$. 

The band of states around 3850MeV, featuring the ground state in the $A_2$ irrep which is safely identified as a $3^{--}$ state,  can be naturally explained as the D-wave triplet. The two very close states above $J/\psi(2S)$ in the $T_1$ channel  are expected to be a $1^{--}$ and a $3^{--}$. The overlaps onto the operator $(\rho \times D^{[2]}_{J=2})^{J=3}$ give
\begin{equation}
\frac{Z_{A_2}(\text{ground state})} {Z_{T_1}(\text{2nd exited state})} = 79(6), \quad \frac{Z_{A_2}(\text{ground state})} {Z_{T_1}(\text{3nd exited state})} = 1.06(1). 
\end{equation}
This shows that the third exited state should be a $3^{--}$ while the second exited state is a $1^{--}$. The ground state in $E$, which is much lower than the other states in this channel, can be identified as a $2^{--}$. The ground state in $T_2$ is associated with the ground state in $E$ by looking at the overlaps onto $(\rho \times D^{[2]}_{J=2})^{J=2}$: 
 \begin{equation}
 \frac{Z_E(\text{ground state})}{Z_{T_2}(\text{ground state})} = 1.02(1).
 \end{equation}
 The first exited state in $T_2$ is associated with the ground state in $A_2$ by the relation of their overlaps onto the $J=3$ operator $(\rho \times D^{[2]}_{J=2})^{J=3}$: 
 \begin{equation}
  \frac{Z_{A_2}(\text{ground state})}{Z_{T_2}(\text{1st exited state})} = 1.04(1).
 \end{equation}    
 In the range above 4000~MeV, we identify three $1^{--}$ states in the $T_1$ channel by looking at the overlap histogram. The state with mass 4189(54)~MeV, indicated in the plot with lighter shading, shows a dominant overlap onto the operator $(\pi \times D^{[2]}_{J=1})^{J=1}$, which is built from a gluonic excitation coupled to a $q\bar{q}$. Therefore, this state is probably not a conventional $q\bar{q}$ state, but could be a hybrid with non-exotic quantum numbers. As we will show below, possible non-exotic hybrids also appear in $0^{-+}$ and $2^{-+}$ at the same energy range. Together with the exotic $1^{-+}$ state, they form a multiplet of near-degenerate hybrid mesons with $J^{PC} = (0 ,1, 2)^{-+}, 1^{--}$ as discussed in Ref.~\cite{Dudek:2011bn}. The other two $1^{--}$ states in $T_1$ channel are possibly radial excitations of the $J/\psi$. 

The overlap histogram of the $A_1$ channel clearly indicates that the ground state is a $4^{--}$. We also find a $4^{--}$ state in each of the $T_1$, $T_2$, and $E$ channels. Their overlaps onto a $J=4$ operator 
$(a_1 \times D^{[3]}_{J_{13}=2, J=3} )^{J=4}$ gives the relations $Z_{A_1}/Z_{T_1} = 1.10(5)$, $Z_{A_1}/Z_{T_2} = 0.93(3)$, 
$Z_{A_1}/Z_{E} = 1.03(3)$, showing that they are different components of a single $4^{--}$ state. In this energy range, we would expect it to be a G-wave.           
\subsection{$J^{+-}$}
\begin{figure}[h]
\centering
\includegraphics*[width=0.7\textwidth]{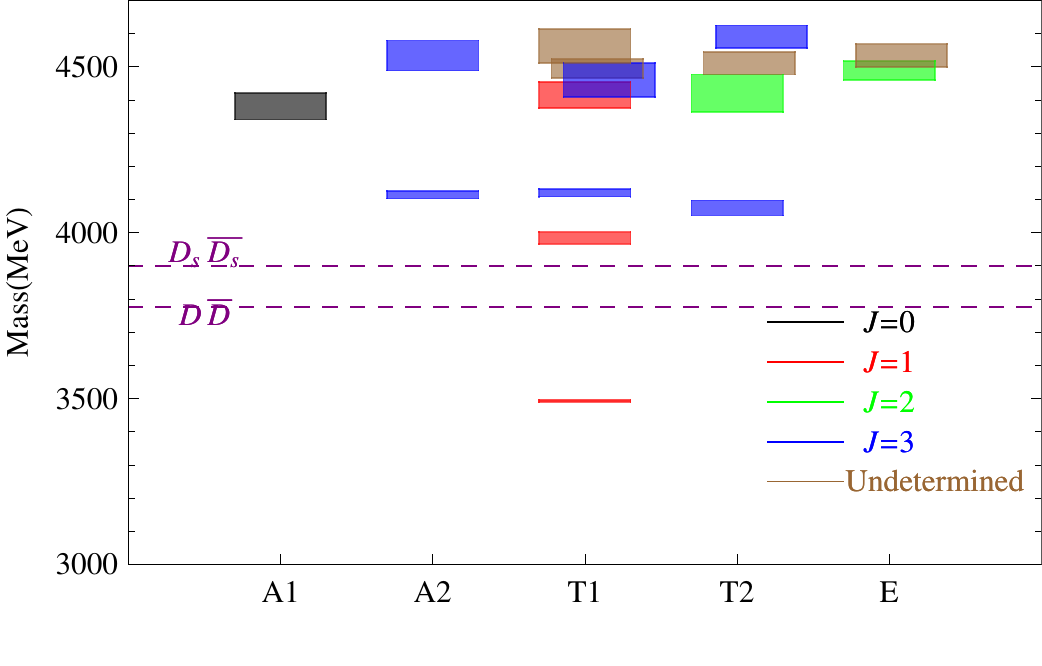}
\caption{\label{Fig:Jpm} The plot shows the masses of the extracted $J^{+-}$ states by lattice irreps. }
\end{figure}

In the $J^{+-}$ sector, the $J$-even states are exotic in the sense that they are not allowed in the quark model.  In Figure~\ref{Fig:Jpm} we show the extracted $J^{+-}$ states by lattice irreps. 
The lightest state in $T_1$ is far lower than the other states and has no partners in other channels. We can safely assign it as the $1^{+-}$, the $h_c$. The set of  states around 4000~MeV are assigned to be a $1^{+-}$, $3^{+-}$ pair.  The ground  state in $A_2$ can be taken as a $3^{+-}$ considering the continuum spin content of $A_2$. The ground state in $T_2$ could be a $2^{+-}$ or a $3^{+-}$. Since $2^{+-}$ is exotic, we expect it be higher than a non-exotic $3^{+-}$ state. The overlap histogram also indicates that the ground state should be a $3^{+-}$. The first and second exited states in $T_1$ channel are assigned as a $1^{+-}$ and a $3^{+-}$ respectively. These assignments are confirmed by comparing the overlaps of the $3^{+-}$ states onto a $J=3$ operator $b_1 \times D^{[2]}_{J=2})^{J=3}$  in $A_2$, $T_1$ and $T_2$:
\begin{equation}
 \frac{Z_{A_2}(\text{ground state})}{Z_{T_1}(\text{2nd exited state})} = 1.01(1), \quad  \frac{Z_{A_2}(\text{ground 
state})}{Z_{T_2}(\text{ground state})} = 1.04(2).
\end{equation}
We assign the ground state in $A_1$ to be an exotic $0^{+-}$ state. The first exited state in $T_2$ are 
associated with the ground state in $E$ by overlap analysis. They complete an exotic $2^{+-}$ state.
\subsection{$J^{-+}$}
\begin{figure}[h]
\centering
\includegraphics*[width=0.7\textwidth]{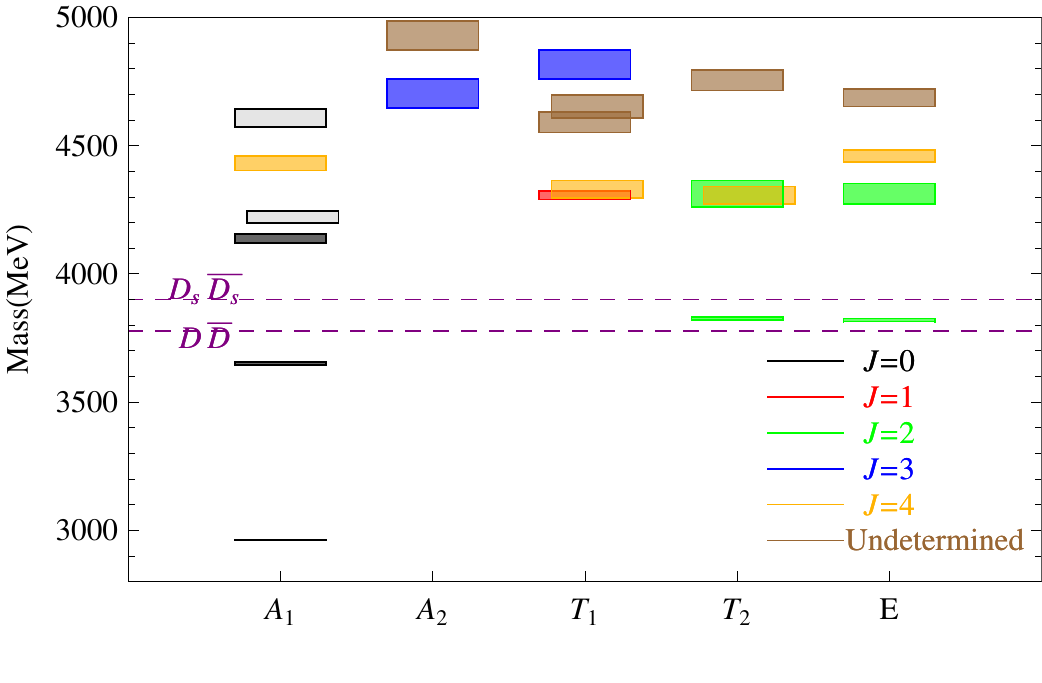}
\caption{\label{Fig:Jmp} The plot shows the masses of the extracted $J^{-+}$ states by lattice irreps.}
\end{figure}

In the $J^{-+}$ sector, the $J$-odd states are exotic. Figure~\ref{Fig:Jmp} shows the extracted $J^{-+}$ states by 
lattice irreps. 
The lowest two states below the $D\bar{D}$ threshold in the $A_1$ channel can be identified as 
$0^{-+}$, they are both S-wave, one is $\eta_c(1S)$ and the other one is $\eta_c(2S)$. The second exited state is also a $0^{-+}$, probably another radially excited S-wave according to the similarity of its overlap histogram with 
the lowest two states. The third exited state has large overlap onto the operator $(\rho \times D^{[2]}_{J=1})^{J=0}$ which contains gluonic excitation so this could be a hybrid state with non-exotic quantum numbers. 

The ground state in $T_2$ and $E$ can be easily identified as a $2^{-+}$ D-wave. 
The first exited states in these two channels are also assigned as $2^{-+}$ and in this energy range could be a radially 
excited D-wave $2^{-+}$ state. 
However, looking carefully at the overlaps, we find that they both have big overlaps onto the operator $(\rho 
\times D^{[2]}_{J=1})^{J=2}$. Furthermore, the ratio of their overlaps onto this operator is $Z_{T_2}/Z_E = 1.2(1)$ 
suggesting that it is probably a $2^{-+}$ hybrid state. 

The ground state in $T_1$ is assigned to be an exotic $1^{-+}$ state. A non-exotic $4^{-+}$ state is located very 
closely. In $A_1$, $T_2$ and $E$ channels, we also find a $4^{-+}$ state. Their overlaps onto the $J=4$ operator 
$(b_1 \times D^{[3]}_{J_{13}=2, J=3} )^{J=4}$ give the relations: 
\begin{equation}
\frac{Z_{A_1}}{Z_{T_1}} = 1.07(4),  \quad \frac{Z_{A_1}}{Z_{T_2}} = 1.09(5),  \quad \frac{Z_{A_1}}{Z_E} = 0.96(3), 
\end{equation}
indicating that they are different components of a $4^{-+}$ state. The discrepancy of the mass in different irreps is possibly a discretisation effect. 

The ground state in $A_2$ could be an exotic $3^{-+}$ state or a non-exotic $6^{-+}$ state. Since we do not use 
operators with overlap onto $J=6$ states, a $3^{-+}$ interpretation is preferable. 
We also find a $3^{-+}$ state in $T_1$ channel, which is associated with the ground state in $A_2$ by overlap analysis. A $3^{-+}$ is also expected in $T_2$ channel to complete this state. However, we don't have convincing evidence to unambiguously 
identify a $3^{-+}$ in $T_2$. 
\section{Summary}\label{Sec:Summary}
We have presented a progress report on our study of the charmonium spectrum using anisotropic lattices and $2+1$ 
fermion flavours. The spin-identification, operator construction and variational analysis used here is crucial for 
an accurate identification of states. In particular we are able to reliably identify spin 4 states and both exotic and 
non-exotic hybrids. 

We have investigated the dependence of the bare fermion anistropy on the quark mass and find that a re-tuning for 
heavy quarks is necessary. The anisotropy measured from the charmonium dispersion relation is then 3.50(2) for 
a target anisotropy of 3.5. It is also interesting to note that the dispersion relation is fully consistent with 
a relativistic interpretation and momenta up to $n=2$ can easily be included in the fit.  

\begin{figure}[h]
\begin{center}
\includegraphics*[width=1.0\textwidth]{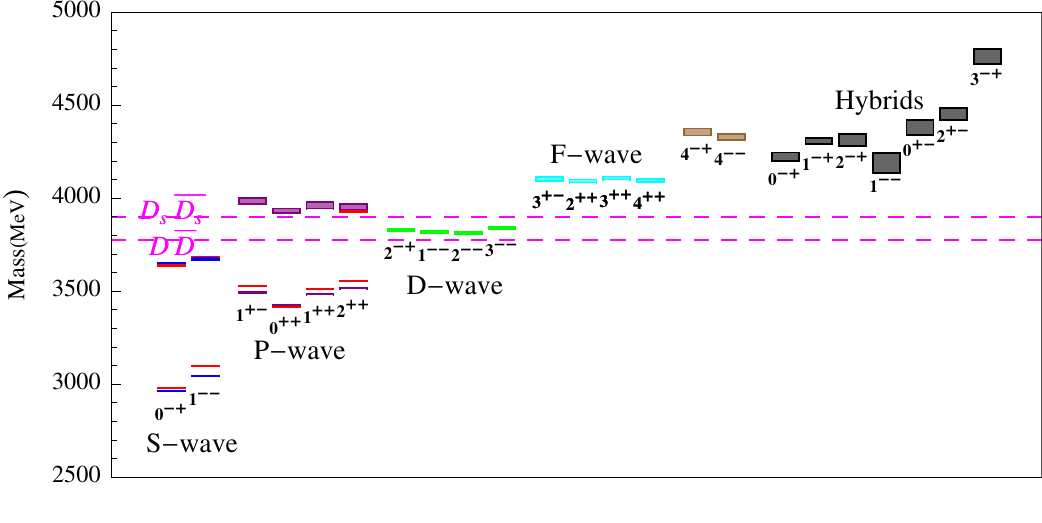}
\end{center}
\caption{\label{Fig:Summary} Summary of the all the charmonium and exotic states. The dashed purple lines 
indicates the $D\bar{D}$ and $D_s\bar{D_s}$ thresholds. The red bars are the experimental values.} 
\end{figure}

Figure~\ref{Fig:Summary} is a summary of the spectrum discussed in the preceeding sections. 
We note that a number of hybrid states, exotic and non-exotic, are determined and we find strong evidence for a charmonium 
``super-multiplet'' as discussed in Ref~\cite{Dudek:2011bn} for light mesons. 
The statistical resolution of our determination of these states, with 96 configurations, is $\sim 17$ MeV. For states below threshold the precision is of the order 1\% or less: $\sim 1$ MeV on the $\eta_c$ and $J/\psi$. 

A complete description of all disconnected effects in the $\eta_c$ is also outlined above. Using distillation this 
calculation is considerably simplified and signals persisting over 5-10 timeslices on just 39 configurations are 
resolved. This ongoing work with higher statistics will be more completely described in a further paper. 

\subsection*{Acknowledgements}
We thank our colleagues in the Hadron Spectrum Collaboration. 
We are grateful for support from the Research Executive Agency (REA) of the European Union under Grant Agreement number
 PITN-GA-2009-238353 (ITN STRONGnet). SR is supported by the Science Foundation Ireland, 
Grant No. 11/RFP.1/PHY/3201. 

\bibliography{Proceeding_2011Latt}
\end{document}